\def\unredoffs{\voffset=-0.5in\hoffset=-0.5in}

  \documentstyle[12pt]{article}
  \unredoffs
\begin{document}
\begin{titlepage}
\begin{flushright}
TIFR/TH/94-23\\
June 1994
\end{flushright}
\begin{center}
{\Large \bf UNIVERSAL CELLULAR AUTOMATA \\
AND CLASS 4}
\end{center}
\bigskip
\renewcommand{\thefootnote}{\fnsymbol{footnote}}
\begin{center}
Avinash Dhar, Porus Lakdawala, Gautam Mandal,
 Spenta R. Wadia\footnote{E-mail: {\tt
adhar/porus/mandal/wadia@theory.tifr.res.in}}\\
Tata Institute of Fundamental Research\\
Homi Bhabha Road, Bombay-400005,INDIA.
\end{center}
\begin{abstract}
  Wolfram has provided a qualitative classification of cellular
  automata(CA) rules according to which, there exits a class of CA
  rules (called Class 4) which exhibit complex pattern formation and
  long-lived dynamical activity (long transients). These properties of
  Class 4 CA's has led to the conjecture that Class 4 rules are
  Universal Turing machines {\it i.e.\/} they are bases for
  computational universality.  We describe an embedding of a ``small''
  universal Turing machine due to Minsky, into a cellular automaton
  rule-table. This produces a collection of $(k=18,r=1)$ cellular
  automata, all of which are computationally universal. However, we
  observe that these rules are distributed amongst the various Wolfram
  classes. More precisely, we show that the identification of the
  Wolfram class depends crucially on the set of initial conditions
  used to simulate the given CA.  This work, among others, indicates
  that a description of complex systems and information dynamics may
  need a new framework for non-equilibrium statistical mechanics.
\end{abstract}
\bigskip
\bigskip
\smallskip
\end{titlepage}

\renewcommand{\thefootnote}{\arabic{footnote}}
\setcounter{footnote}{0}

\section*{\protect\normalsize\bf 1. Introduction}

The modern computer epitomises the pinnacle of machine complexity.
More than half a century ago A. Turing had formalised the notion of
the effective procedure or algorithm, through the
introduction of what is now well-known as the Turing Machine.
It is the Universal Turing Machine that is the formal analogue
of our general-purpose computer.  Put simply, a Universal Turing
machine is an automaton which, when given suitable instructions, can
do anything that can be done by automata at all.

It was von Neumann\cite{neumann} who first introduced the cellular
automaton. The von Neumann automaton, is a self-reproducing unit and
is equivalent, by construction, to a Universal Turing Machine.

More recently, cellular automata have become paradigms for
complex ``life-like'' systems. Among other things, they provide
an ideal substratum to simulate artificial, biological
environments\cite{lang}. In this sense, they might provide
a suitable, abstract setting for discussions of biological
complexity.

In this paper we investigate the connection between
a certain class of cellular automata (technically called
Class 4) and the notion of universal computation. Our main result
is that the CA rule-based classification proposed by Wolfram seems
to be inadequate. Any quantitative classification scheme would
have to take into account the space of initial configurations.
We believe that this phenomenon may have implications for
non-equilibrium statistical mechanics.

Our paper is organised as follows. In section 2 we collect for
completeness the definiton of a cellular automaton and the Wolfram
classes. In section 3, we briefly review the studies of Langton {\it
  et al\/} on the connection between cellular automata and phase
transitions in statistical mechanics. Section 4 contains a brief
review of universal Turing machines as well the construction of
Minsky's ``small UTM''. Section 5 contains the main result. We then
conclude with some observations and remarks. In the appendix we have
collected some of the well-known definitions from information theory
that we will need in the discussion.

\section*{\protect\normalsize\bf 2. The Wolfram Classes}

Cellular automata(CA) are discrete (both in space and time) dynamical
systems. More formally, consider variables sitting at the sites of a
one-dimensional lattice. The variables take values from a finite set
$S$\@. The evolution of the CA proceeds through discrete time-steps
by a local rule, which is specified by the function
\begin{equation}
x_{i}^{t+1} = f(x_{i-r}^{t},\ldots ,x_{i}^{t},\ldots ,x_{i+r}^{t})
\end{equation}
where $x_{i}^{t}$ denotes the value of the variable at the lattice-site
$i$ at time $t$. A CA whose lattice variables take one of $k$
possible values and whose evolution rule depends on at most $r$ neighbours
of a given site is called a $(k,r)$ CA. Moreover, one isolates a special
state $s$ in $S$, called the quiescent or stable state,
as the one that is preserved by the evolution {\it i.e.\/}
$f(s, \ldots ,s) = s$. All CA's considered here have a stable state.
The function $f$ specifying the CA rule is conventionally called
the rule-table, the ($2r+1$)-tuple ($x_{i-r}^{t}, \ldots ,x_{i+r}^{t}$)
is called a template.

In a study initiated by Wolfram\cite{wolf1,wolf2}, he sought to
classify the rule-space of $(k,r)$ CA's on the basis of the
video-displays which are obtained by the evolution of cellular
automata on a computer. He isolated four qualitative classes
of behaviour (now widely known as the Wolfram classes), namely,
\begin{enumerate}
\item Class 1 --- evolution leads to a homogeneous (stable) configuration.
\item Class 2 --- evolution leads to periodically repeating patterns.
\item Class 3 --- evolution leads to ``chaotic'' patterns.
\item Class 4 --- evolution leads to complex patterns, generated by mobile
interacting structures which are relatively long lived.
\end{enumerate}
 It is important to note that this classification is based on the
  evolution of a given rule from a randomly chosen initial
  configuration. Wolfram, further  conjectured that the rules
  belonging to Class 4 possess the capacity to perform universal
  computation\footnote{For a precise definition of the notion
of universal computation see section 4.}\cite{wolf2}.

\section*{\protect\normalsize\bf 3. Phase transitions and Class 4}

There have been numerous attempts in the past few years either to
provide a quantitative basis for Wolfram's classification or to
propose alternative classification schemes. We note two such attempts.
The first is due to Cullick II and Yu\cite{cullick1,cullick2}, who
have tried to use a computation theoretic approach as a basis for
classification. It is a remarkable fact that their classification
coincides with Wolfram's for the ($k=2,r=2$) totalistic
CA's\cite{cullick1}. An important fact, which is of relevance to our
discussion, is that in this classification all CA's which are
  computationally universal belong to Class 4\footnote{In this
  classification, the different classes are statements about all
  initial conditions for the given rule.}.

The other attempt at trying to provide a quantitative understanding of
the Wolfram classes that we would like to discuss is due to Langton and
co-workers\cite{lang1,lang2}, who made a statistical study of the rule
spaces of various $(k,r)$ CA rules.  They used various information
theoretic quantities like the Shannon entropy and mutual
information\footnote{For definitions of Shannon entropy and mutual
information see appendix A.}. In what follows, we briefly describe this
work.  For details, we refer the reader to the original sources.

{}For a given rule-table, Langton introduced a parameter, $\lambda$,
defined as the fraction of templates in the rule-table which are
mapped to a non-quiescent state. It is intuitively quite clear that
rules with a low value of $\lambda$ (close to $0$) belong either to
Class 1 or 2, while those with a high value of $\lambda$ (close to
$1$) belong to Class 3. The ``complex'' Class 4 rules are expected to
occur at intermediate values of the $\lambda$ parameter.  Consider a
collection of randomly generated rules, one at each value of
$\lambda$, spanning the interval $[0,1]$. We will call such a
collection a lambda-string.  If we now think of the rule-space as an
abstract space where each point represents a CA rule, then a
particular lambda-string might be thought of as a curve through the
rule-space. What Langton noticed was that, while along most curves (or
equivalently, lambda-strings) the transition from Class 2 rules to
Class 3 rules was discontinuous (sharp), there did occur some curves
in the rule-space along which the transition from Class 2 to Class 3
was smooth. These curves in-fact ``passed through'' Class 4 rules. In
more quantitative terms, the Shannon entropy, when calculated over
lambda-strings of the former type, showed a jump at some value of
$\lambda$ (which depended on the chosen string), from values quite
close to $0$ to values near $1$. The value of the mutual information
at different points on such curves did not differ substantially from
$0$.  However, for lambda-strings of the latter type, the entropy
showed a relatively smooth transition from values near $0$ to values
near $1$. The mutual information, on the other hand, showed a sharp
peak for these lambda-strings.  Langton concluded that the structure
of the rule-space appeared to be as follows:

There is an ``ordered phase'' of Class 1 and 2 CA rules, separated,in
general, from the ``disordered or chaotic phase'' of Class 3 rules by a
``first order'' transition. However, in the vicinity of this phase
boundary lie pockets of ``complex'' Class 4 rules. If the rule-space is
traversed across these pockets of complex rules, the order-to-chaos transition
is a ``second order'' transition.

The order of the phase transition is to be understood, by treating the
Shannon entropy in analogy with the usual entropy in statistical
mechanics and the mutual information, with the derivative of the
entropy.

\section*{\protect\normalsize\bf 4. Minsky's ``small'' UTM}

We briefly review here the definition of a Turing machine.
{}For details, the reader may refer any standard text on computation
theory like \cite{minsky,hopcroft}.

A Turing machine(TM) consists of a ``head'' which moves along
 an infinite ``tape'' consisting of cells. The cells on the tape can
 each carry a symbol from a finite set $Q$. There is a special
 symbol in $Q$ called the blank. Initially, all except for a
 finite number of cells on the tape carry the blank. The head of the
 TM, on the other hand, can exist at each instant in one of a finite
 number of states chosen from some finite set $T$. $T$ contains
 a special state called the start state. Initially, the head of
 the TM is in the start state. At a given instant of time, the
 head resides at a particular cell on the tape. Depending on the
 tape-symbol that is ``read'' by the head and also on the particular
 state that the head is currently in, the following transformations
 are allowed:
\begin{enumerate}
\item[1] The present tape-symbol may or may not be altered to a new symbol.
\item[2] The present head-state  may or may not be altered to a new state.
\item[3] The head will move one cell either to its right or left, or else
the TM will halt.
\end{enumerate}
The definition of a particular TM consists in specifying $Q$, $T$
and the state-symbol transition-table.

As was mentioned before, the tape of the TM would initially have all
cells ``blank'' except for a finite number. These non-blank cells can
be thought of as encoding the ``program'' or ``algorithm'' which
controls the evolution of the TM. A Universal Turing Machine
(UTM) is defined to be a TM which can simulate any other TM, if
supplied with an appropriate program which encodes the description of
the TM to be simulated.

We now give the description of a 4-symbol,7-state UTM due to
Minsky\cite{minsky}.  The set of symbols is
$Q=\{q_{0}$,$q_{1}$,$q_{2}$,$q_{3}\}$ and the the set of states is
$T=\{t_{0}$,$t_{1}$,$t_{2}$,$t_{3}$,$t_{4}$,$t_{5}$,$t_{6}\}$. $q_{0}$
denotes the blank symbol, while $t_{1}$ is the start
state. The transition-table for this TM is given below (see Table
\ref{utm}). If the combination of the tape-symbol and the head-state
at a particular instant is $(q_{i},t_{k})$, the the corresponding
entry in the table gives the appropriate transformation for the TM.
Here $r$ denotes ``move to the right'', $l$ denotes ``move to the
left'', while $h$ denotes ``halt''.  The proof that this TM is indeed
computationally universal can be found in \cite{minsky}.

\begin{table}
\caption{The symbol-state transition-table of Minsky's small UTM.
The entry in the table corresponds to the transformation that the
TM would perform if the head is currently in the state $t_{i}$
and is reading the tape-cell containing the symbol $q_{j}$.
$l$ denotes a left-move, $r$ denotes a right-move, $h$ denotes halt.}

\begin{center}
$
\begin{array}{|c|ccccccc|}   \hline
&&&&&&&\\
     & t_{0}          & t_{1}          & t_{2}          & t_{3}          &
        t_{4}          & t_{5}          & t_{6}   \\
&&&&&&&\\ \cline{2-8}
&&&&&&&\\
{}~~q_{0}~~& (q_{0},t_{0},l)& (q_{2},t_{1},r)& (q_{0},t_{2},h)&
(q_{2},t_{4},r)&
        (q_{2},t_{2},l)& (q_{3},t_{2},l)& (q_{2},t_{5},r)  \\
&&&&&&&\\
q_{1}& (q_{1},t_{1},l)& (q_{3},t_{1},r)& (q_{3},t_{2},l)& (q_{1},t_{6},l)&
        (q_{3},t_{4},r)& (q_{3},t_{5},r)& (q_{1},t_{6},r)  \\
&&&&&&&\\
q_{2}& (q_{0},t_{0},l)& (q_{0},t_{0},l)& (q_{2},t_{2},l)& (q_{2},t_{3},l)&
        (q_{2},t_{4},r)& (q_{2},t_{5},r)& (q_{0},t_{6},r)  \\
&&&&&&&\\
q_{3}& (q_{1},t_{0},l)& (q_{2},t_{5},r)& (q_{1},t_{3},l)& (q_{1},t_{3},l)&
        (q_{1},t_{4},r)& (q_{1},t_{5},r)& (q_{0},t_{1},r)  \\
&&&&&&& \\ \hline
\end{array}
$
\end{center}
\label{utm}
\end{table}

\section*{\protect\normalsize\bf 5. Universal Computation and Class 4}

Universal computation is, by definition, the domain of performance of
a Universal Turing Machine. A Universal Cellular Automaton(UCA) is one
which simulates, at every time step, a Universal Turing Machine(UTM).
A CA which has been proven to be universal is the well known Game of
Life, due to Conway\cite{conway}.  In general, it is difficult to
decide whether a given CA is a UCA. It is in fact easier, to construct
CA's which are universal.  Small UCA's are obtained by ``embedding'' a
small universal Turing machine into a CA rule. Of these methods, which
by now are quite well known, we describe one a little later.

{}From the discussion in sections 2 and 3, it appears that the Class 4 CA
rules bring together two rather disparate looking themes --- that of
phase transitions from statistical mechanics and universal computation
from computation theory. However, at present it can hardly be said
that the connection between these two themes is clear.

The aim of the present investigation is to provide a somewhat better
understanding of the connection between Class 4 CA's and universal
computation. In a sense this work may be thought of as an effort
in a direction, complementary, to that of Langton {\it et al\/} in
addressing the above question.

 The idea employed in the present work is quite simple and
is as follows:

 Using a well-known construction of a ``small'' UTM due to
  Minsky\cite{minsky} we construct a rather large collection of
  $(k=18,r=1)$ CA rules all of which are universal. We then
  perform a Wolfram classification of the CA rules within this
  subspace of the rule-space.  Surprisingly, one finds that even
  within the subspace of UCA's there are rules which seem to belong to
  each of the Wolfram classes (except Class 1).

  We now describe this result in greater detail. First we demonstrate a
  simple and very well-known way\cite{smith} of embedding any TM into
  a CA.

An embedding of a TM into a CA requires:
\begin{enumerate}
\item[1] The specification of a mapping between the states and symbols
of the TM and the states of the CA, which would allow us, at every
time step, to transform a TM tape-head configuration into the
corresponding CA configuration.
\item[2] The specification of the rule-table of the CA which is
consistent with the state-symbol transition-table of the TM with
respect to the above mapping.
\end{enumerate}

To construct the embedding, think of the CA lattice as the tape of the
TM. The lattice variables should thus carry the tape-symbols of the
TM. Moreover, they should also carry information about the position
and the state of the TM head. This can be done by introducing CA
states corresponding to the different head-states of the TM, along
with the information that the head is currently reading the cell to
either its immediate right or left, specified by the symbols ${L,R}$.
The map required in (1) above can now be chosen as follows: The
  set of states $S$ of the CA consists of tape-symbols as well as
  ordered pairs of the head-states and $L$ or $R$ {\it i.e.\/} $S = Q
  \cup (T\times\{L,R\})$. The number of CA states is evidently $|S|\,
=\,|Q|\,+\,2|T|$.  With this map between the CA states and the TM
symbol/states, it is not hard to construct a CA rule-table, with
nearest-neighbour interaction, which simulates the transition-table of
the TM \footnote{ We can also construct an embedding in which the CA
  states are just the TM states and symbols, along with the
  prescription that the ``head'' variable always reads the cell which,
  say, is to its immediate right. This leads to a CA rule-table with a
  next-nearest-neighbour interaction.}.

To clarify the procedure described rather abstractly above, we
reconstruct some of the details now with reference to Minsky's UTM
(see section 4).  In this case the set of states for the CA would be
$\{q_{i},(t_{j},L), (t_{j},R)| i=0,\ldots ,3, ~j=0,\ldots ,6\}$. We
now give two examples of the evolution of the CA lattice-configuration
in a single time-step.  From these examples it is clear how the
rule-table for the CA can be constructed.

In the first example the UTM at the present time step has symbols
$\ldots ~q_{0} ~q_{2} ~q_{1} ~q_{2} \ldots$ inscribed on the tape
and the head is in the state $t_{4}$ pointing at the tape cell
containing $q_{1}$. From the appropriate entry in Table 1 the UTM
evolves by changing the tape-cell from $q{1}$ to $q_{3}$ and the
head moves a step to the right without changing its state. In terms
of the CA this evolution could be represented as follows:

\begin{center}
$
\begin{array}{ccccccc}
  \ldots & q_{0} & q_{2} & (t_{4},R) &   q_{1}    & q_{2} & \ldots \\
  \ldots & q_{0} & q_{2} &   q_{3}   & (t_{4},R)  & q_{2} & \ldots \\
\end{array}
$
\end{center}

In the second example the UTM performs the same evolution as in
previous example. However in terms of the CA it could also be
represented as follows:

\begin{center}
$
\begin{array}{ccccccc}
  \ldots & q_{0} & q_{2} & q_{1} & (t_{4},L)  & q_{2} & \ldots \\
  \ldots & q_{0} & q_{2} & q_{3} & (t_{4},R)  & q_{2} & \ldots \\
\end{array}
$
\end{center}

Note however, that this ambiguity would come into play at only the
initial time-step and could be removed by demanding that only one of
the configurations, say the one in the first example, can be a legal
initial configuration. After making this demand, the rest of the CA
(or equivalently the UTM) evolution is completely unambiguous.

As a result of the prescription described above, one obtains a large
class of ($k=18,r=1$) CA's, all of which are computationally
universal. The large class of UCA's obtained is accounted for by the
fact that the embedding does not fix all the ($2r+1$)-tuples
($x_{i-r}^{t}, \ldots ,x_{i+r}^{t}$) of the rule-table to a unique
value. Since the UTM contains only a single ``head'', the
($2r+1$)-tuples which contain only a single head-state are uniquely
determined by the definition of the UTM. However, CA
lattice-configurations which contain more than one head-state are
perfectly legal as far as the cellular automaton is concerned.
{}For example the following evolution is perfectly legal in terms
of the CA:

\begin{center}
$
\begin{array}{ccccccc}
  \ldots & (t_{6},R) &   q_{2}   & q_{1} & (t_{4},L)  & q_{2} & \ldots \\
  \ldots &   q_{0}   & (t_{6},R) & q_{3} & (t_{4},R)  & q_{2} & \ldots \\
\end{array}
$
\end{center}

Although such configurations would make no sense in terms of the
underlying Turing machine, they have to be considered while performing
the Wolfram classification. It is of importance to mention that for
initial configurations which contain a single head-state, all rules
within the space of UCA's that we are considering have the same
evolution. Within the space of CA lattice-configurations those which
contain two or more head-states form an overwhelming majority, and if
one is to select a (few) random initial configuration(s) as a basis of
classification, then it would invariably be one of these.

 In other words, the subspace of CA lattice-configurations which
  governs the behaviour of any of the UCA's as a UTM is quite distinct
  from the subspace of configurations which determines the Wolfram
  class, to which the UCA belongs.

  In Figures 1 and 2, we have shown the variation of the Shannon
  entropy and the mutual information over a (generic) lambda-string
  through the subspace of UCA's. The decay of mutual information to
  $0$ on both sides of a peak value, suggests the wide variation in
  complexity of the rules associated with the lambda-string. The
  video-displays shown in Figures 3--5, corroborate this fact. We
  remark that we have defined $\lambda$ as the fraction of templates
  of the rule-table, not fixed by the definition of the UTM, that are
  mapped onto the non-quiescent state.

\section*{\protect\normalsize\bf 6. Conclusions and Remarks}

The main observation that emerges from our investigations is the
importance of initial conditions for any study of complex systems.  As
far as CA's are concerned, any quantitative classification of the
rule-space must also, perforce, be a statement about the space of
initial conditions. We draw the reader's attention, once again, to the
classification due to Cullick II and Yu\cite{cullick1}, where the
definitions of the classes are statements about all initial
conditions.  Whether the agreement of these classes, with that of
Wolfram's, for the $(k=2,r=2)$ totalistic rules is a mere coincidence,
or a necessity for small rule-spaces remains unclear. Our analysis
shows that the subspace of $(k=18,r=1)$ rules that we have considered,
which would belong entirely to Class 4 of the Cullick-Yu scheme, by
virtue of their being computationally universal, are actually
distributed amongst the various Wolfram classes.

Complex systems are invariably studied from one of two different
points of emphasis. One is the view arising out of statistical
mechanics and dynamical systems, where the emphasis is on the
properties of the system observed at large times {\it i.e.} the
steady- state properties of the system. The second view is the one
that arises from formal studies of the computational complexity
classes.  Here the emphasis is on the behaviour of the system as a
function of the input. We feel that a new framework of statistical
mechanics which incorporates the second point of view, is required
to provide a better definition for a study of complex dynamical
systems.

The observations made above, seem to suggest an avenue for further
exploration. One might be tempted to consider a scenario
in which the classes are not well demarcated regions of the rule-space,
but rather, are sets with fuzzy boundaries. The measure of fuzziness
attributed to the set, could depend on the proportion of the space of
initial configurations on which the rules (contained in the set) evolve
in a complex manner.


\appendix

\section*{\protect\normalsize\bf Appendix A. Some Definitions from
 Information Theory}

In this appendix we give the definitions of information theoretic
quantities, like the Shannon entropy and mutual information, which we
have used as measures of pattern complexity. For details
see, for example, \cite{klir} and references therein.

Consider a probability measure $\wp_{X}$ on a finite set $X$. We have
for any ``event'' $x \in X$, the associated probability $p(x)$.
The uncertainity associated with the event $x$ is defined to
be $log_{2}(p(x))$. This is a natural definition, if we require
that the uncertainity for a pair of independent events, be the sum
of their individual uncertainities. The information gained
when event $x$ is realised in an experiment is defined as
$-log_{2}(p(x))$ {\it i.e.} the negative of the uncertainity
associated with $x$\footnote{The base of the logarithm, specifies the
units in which the information is measured. Conventionally, the base
is chosen to be $2$ and the information is then said to be measured in
bits.}.

The Shannon entropy $H(X)$ is defined as the average information content
of the distribution {\it i.e.}
\begin{equation}
H(X) = - \sum_{x\in X} p(x)\log_{2}(p(x))
\end{equation}

It can be shown that the Shannon entropy is the unique measure
of uncertainity that can be defined, satisfying some rather
general requirements\cite{klir}.

Consider two finite sets $X$ and $Y$, and a joint probability
distribution $\wp_{X\times Y}$ on $X\times Y$. We can thus define the
corresponding Shannon entropy which we denote by $H(X,Y)$.  Moreover,
using the marginal probability distributions on $X$ and $Y$, induced
by this joint distribution, we can define the individual
Shannon entropies $H(X)$ and $H(Y)$ as before.

The mutual information $M(X,Y)$ (also called the information
transmission) is defined as
\begin{equation}
M(X,Y) = H(X) + H(Y) - H(X,Y)
\end{equation}

It is possible to show that $H(X,Y) \leq H(X) + H(Y)$, so that
$M(X,Y) \geq 0$. The equality holds in the case when $X$ and $Y$
are probabilistically independent. Thus the mutual information
measures correlation with respect to the joint probability, of
the sets $X$ and $Y$.

The quantities that we have discussed above can easily be calculated
for the video-displays that arise on the evolution of a CA on a
computer. We define a cell as a subset of the CA lattice-sites.
We choose for the set $X$ the possible configurations of this cell.
What remains is to find a natural probability distribution on $X$.
This is easily done by counting the frequency with which the various
configurations for the given cell occur in time, when we evolve the CA from,
say, a (few) random initial lattice-configuration(s). This
distribution can then be used to calculate the Shannon entropy of
this cell. We could also calculate the mutual information between
two different cells. The resultant quantities can be thought of
as measures of complexity of the patterns that are generated
by the given rule.

\end{document}